\documentclass[12pt]{article}
\usepackage[utf8]{inputenc}
\usepackage{authblk}
\usepackage{setspace}
\usepackage[margin=1 in]{geometry}
\usepackage{graphicx}
\graphicspath{ {./figures-new/} }
\usepackage{subcaption}
\usepackage{amsmath}
\usepackage{svg}
\usepackage{wrapfig}
\usepackage{float}
\usepackage{booktabs} 

\usepackage{caption}
\usepackage[font=small,labelfont=bf]{caption}

\usepackage{hyperref}

\usepackage[style=nature, 
citestyle=numeric-comp,
sorting=none]{biblatex}
\title{Stacked Generative Machine Learning Models for Fast Approximations of Steady-State Navier-Stokes Equations}

\author[1]{Shen Wang}
\author[1]{Mehdi Nikfar}
\author[2*]{Joshua C. Agar}
\author[1,3*]{Yaling Liu}

\affil[1]{\small Department of Mechanical Engineering and Mechanics, Lehigh University, Bethlehem, PA, USA.}
\affil[2]{Department of Materials Science and Engineering, Lehigh University, Bethlehem, PA, USA}
\affil[3]{Department of Bioengineering, Lehigh University, Bethlehem, PA, USA}
\affil[*]{jca318@lehigh.edu}
\affil[*]{yal310@lehigh.edu}

\date{}

\providecommand{\keywords}[1]
{
  \small	
    \textbf{Keywords: } #1
}

\begin{document}

\maketitle

\begin{abstract}
Computational fluid dynamics (CFD) simulations are broadly applied in engineering and physics. A standard description of fluid dynamics requires solving the Navier-Stokes (N-S) equations in different flow regimes. However, applications of CFD simulations are computationally-limited by the availability, speed, and parallelism of high-performance computing. To improve computational efficiency, machine learning techniques have been used to create accelerated data-driven approximations for CFD. A majority of such approaches rely on large labeled CFD datasets that are expensive to obtain at the scale necessary to build robust data-driven models. We develop a weakly-supervised approach to solve the steady-state N-S equations under various boundary conditions, using a multi-channel input with boundary and geometric conditions. We achieve state-of-the-art results without any labeled simulation data, but using a custom data-driven and physics-informed loss function by using  and small-scale solutions to prime the model to solve the N-S equations. To improve the resolution and predictability, we train stacked models of increasing complexity generating the numerical solutions for N-S equations. Without expensive computations, our model achieves high predictability with a variety of obstacles and boundary conditions. Given its high flexibility, the model can generate a solution on a $64 \times 64$ domain within 5 $ms$ on a regular desktop computer which is 1000 times faster than a regular CFD solver. Translation of interactive CFD simulation on local consumer computing hardware enables new applications in real-time predictions on the internet of things devices where data transfer is prohibitive and can increase the scale, speed, and computational cost of boundary-value fluid problems.

\end{abstract}

\section*{Introduction}
Many fluid problems are governed by nonlinear partial differential equations (PDE) based on the N-S equations. Numerical simulations of fluid dynamics require solving Navier-Stokes(N-S) equations in the discretized form spatially and temporally. Various methods such as finite difference method (FDM), finite volume method (FVM) \autocite{Ferziger1999-xt,Versteeg2007-ej} Lattice-Boltzmann method (LBM)\autocite{Chen1992-fe,Nikfar2020-el,Tan2016-pa}, finite element method (FEM)\autocite{Girault1981-iu,Reddy2018-rj,Nikfar2016-pf} can solve the N-S equations. These methods can be computational and memory prohibitive if high-resolution meshes are required. This is further complicated by difficulties in determining the proper computational grids\autocite{Sengupta1988-vd}. Although commercial software relieves some of the burdens of these preprocessing tasks, they still require knowledge of computational fluid dynamics (CFD) such as familiarity with governing equations, choosing proper flow solvers, upwind schemes, and turbulent models \autocite{Duriez2016-gj} and have minimal impact on the computational cost. Furthermore, regardless of computational similarity, when the boundary condition or geometry domain are changed even subtlety, simulations must be reproduced from scratch. \\

There are growing applications in video game engines, ocean current or hurricane forecast, oil spill or fire smoke spreading prediction, and porous media flow etc., where fast automated solutions to the N-S equation are required for physics-informed dynamic control systems and reinforcement learning. Current commercial software is ill-suited for fully-automated applications. Automating and accelerating real-time fluid simulation can enable new applications where the speed, energy, and computational cost are mission-critical \autocite{Zuo2009-rf,Stam2003-pu, Omerdic2007-vl, Biswas2010-bj,Salamonowicz2021-pe, Adler1990-lm}. Therefore, there has been growing interest in developing precise and coherent reduced-order models capable of expressing flow characteristics.\\ 

Machine learning (ML), and especially deep learning (DL) has achieved remarkable successes in computational mechanics by serving as an approximation for dynamic spatiotemporal systems \autocite{Zhang2019-ey,Yang2016-lx,Tompson2017-jt,Raissi2019-mn,Habibi2020-jj}. DL models generally are constructed as overparameterized models that can serve as data-driven approximations between the input and the target. By adding “damaging mechanisms” in the form of regularization, a loss function can be optimized to generalize within the distribution of training data. \\

In a purely data-driven approach, the DL model is utilized as a “black box” to map the input to the output. Generating a comprehensive dataset to train robust DL models, however, is computationally expensive and requires a detailed understanding of the statistical distribution of the training dataset \autocite{Rao2020-oy}, and it is hard to achieve robustness when models are overparameterized and training data is limited \autocite{Qi2020-qk}. Recently, physics-informed neural networks (PINNs) have been used to add physics constraints that improve the generalizability of DL models.  For example, physical laws, including PDEs, initial and boundary conditions are explicitly embedded into the DL model. PINNs use physics as a parsimonious regularizer that enhances the robustness and interpretability of the DL model\autocite{Rao2020-oy}. Trained PINNs can predict solutions of PDEs with different types of boundary and initial conditions as an alternative solver that still requires extensive computation to generate the training data \autocite{Lu2021-do}. Trained models are only valid within the distribution of the training data set and thus are not applicable when the geometry, boundary, and initial conditions are changed. Furthermore, once a model is trained it cannot be easily adjusted for system-level discrepancies (e.g., changes in friction\autocite{Kaheman2019-ay}). Therefore, these networks can only accelerate very narrowly defined problems. \\

Several groups have recently tried to predict the results of different PDEs by building PINNs\autocite{Tompson2017-jt,Raissi2017-sh,Raissi2019-mn,Raissi2019-sl,Lu2021-do,Sharma2018-sq,Sun2020-va,Lu2019-lf}. For example, a weakly-supervised algorithm was capable of solving Laplace’s equations \autocite{Sharma2018-sq}. In this work, a fully convolutional encoder-decoder network in a U-Net architecture \autocite{Ronneberger2015-ls} is used to predict the solution. This work combines physics-informed loss with a generative model \autocite{Farimani2017-bh,Thuerey2020-ik,Werhahn2019-eu} that keeps the intrinsic relations between neighboring nodal points. Without access to any solved simulation data, the trained model could predict the solution with different boundary conditions. However, the method was only applied to the heat equation with Dirichlet boundary conditions. N-S equations are second-order nonlinear PDEs containing multiple equations and variables with convective, pressure, and viscous terms, which are much more complicated than heat equations. The heat equation is similar to just the viscous term in N-S equations. Therefore, this method would not be applicable to the general N-S equations. \\

Here, we focus on solving the N-S equations in two-dimensional space. We develop a weakly-supervised method that considers complicated momentum and continuity equations, pressure field, and velocities in both directions. The steady solutions of flow problems governed by N-S equations are generated in 5 \textit{ms} without computed CFD results. To do this, we conducted warm-up initialization by the pre-running iterative solver or coarse solutions. Using convolutional U-Nets, we accurately approximate the solution of steady N-S equations given different boundary conditions and internal obstacles by training stacked models. To improve the performance we impose physics-informed, data-driven constraints on the loss function, represented by the residue of the equations and the differences between the output and the known values on the boundaries, respectively. We validate the models by comparing the results to the FDM solutions. We achieve solutions with a root mean square error of 0.04 for velocity fields compared to the ground truth FDM simulations while requiring 1000 times less computation time.  \\

\section*{Results}
\begin{figure}[H]
    \centering
    \includegraphics[width = \textwidth]{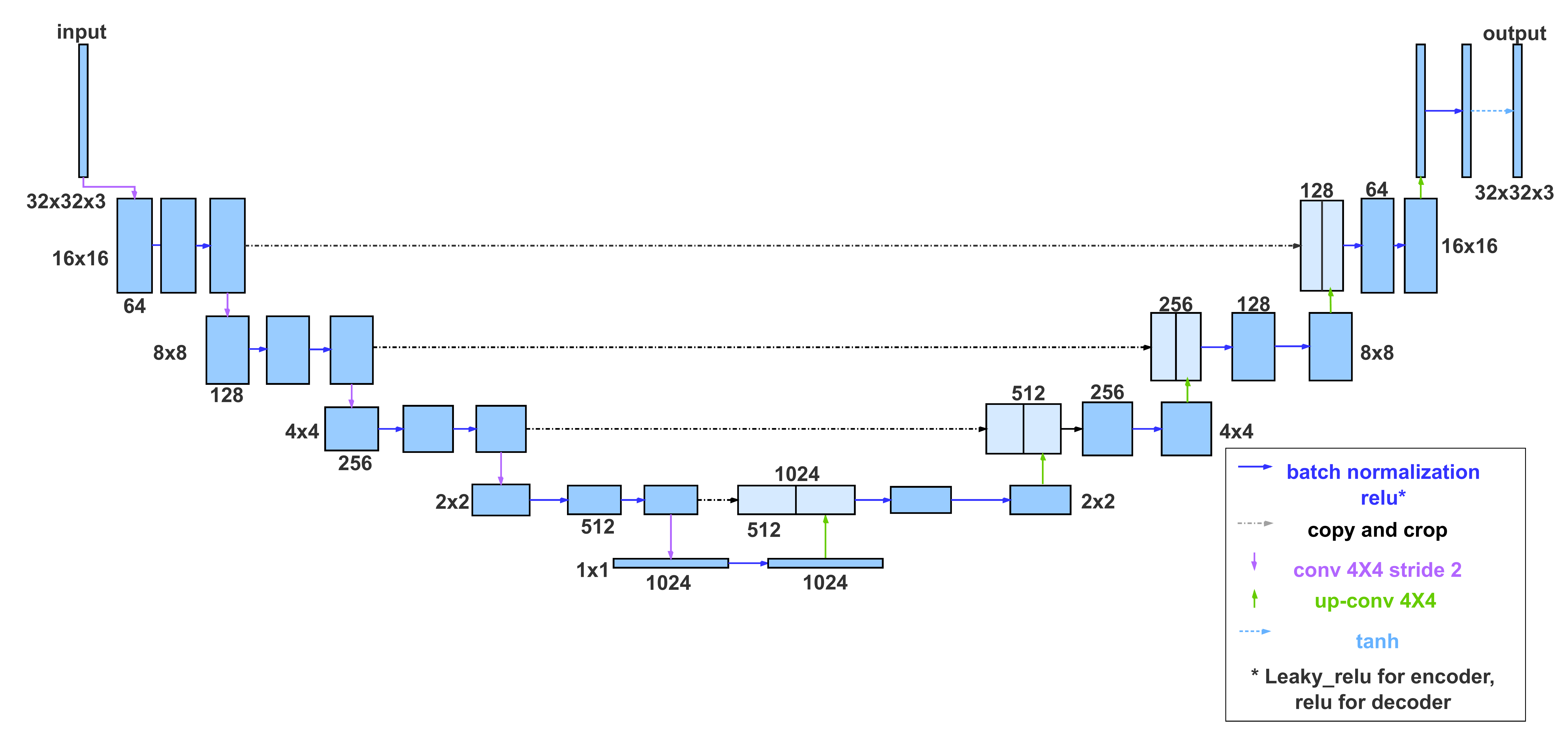}
    \caption{Schematic representation of the architecture of the model. The input of the data can be sized differently. In this figure, the example input domain has a size of $(32 \times 32 \times 3)$ for width, height, and number of channels (depth), respectively. The number of layers in the encoder and decoder is both equal to $\log_2 a$, where $a$ is the dimension of the square matrix (in the shown figure, 5 layers for both the encoder and the decoder). The width and height of the data increase in the encoder and decrease until the original size is recovered in the decoder. The same row of data representation has the same dimension in width and height annotated at left or right, and the same depth except annotated at top or bottom specifically. 
}
    \label{fig:unet}
\end{figure}

To approximate solutions to the N-S equation with boundary conditions, we designed a neural network architecture, the convolutional U-Net (Figure \ref{fig:unet}). The designed U-Nets are a modified autoencoder structure. Autoencoders consist of an encoder that learns a compact representation of data and a decoder that reconstructs a target of similar dimensionality from the compact representation (Figure \ref{fig:unet}). U-Nets include message passing from the encoder to the decoder to assist in reconstructing high-resolution representations. U-Nets have been used for image reconstruction from the input with given constraints in different fields  \autocite{Feng2020-kv,Hu2019-zh,Souza2019-xh}. Generative models have been used for DL based CFD simulations by U-Nets  \autocite{Farimani2017-bh,Takbiri-Borujeni2019-lb,Thuerey2020-ik} and generative adversarial networks (GANs) \autocite{Farimani2017-bh, Werhahn2019-eu}. The designed U-Net is constructed from convolutional layers that take the input and compress the information until reaching the bottleneck layer. The output from the encoder is then passed to the decoder to reconstruct the original size of the input and to map the output to $[-1, 1]$. The output from the final layer is scaled to match the expected magnitude for each channel, which becomes the final output of the model.\\

Optimization based on purely data-driven loss functions resulted in costs for preparing the numerical solutions as labeled data \autocite{Thuerey2020-ik, Takbiri-Borujeni2019-lb}. To improve the model performance with minimal low-cost computationally-simulated data, we constructed a custom loss function that combined the physics informed loss $L_{phy}$ that represents the physics equations and constraints defined by the residue of the computational operations, and the data-driven loss $L_b$ that represents the direct comparisons with the boundary conditions to constrain the model. The physical model used in this study is based on two-dimensional incompressible fluid dynamics constrained by N-S equations. The governing equations can be rewritten as equations \ref{eqn:ns1} to \ref{eqn:ns3} :

\begin{equation}\label{eqn:ns1}
    \frac{\partial u}{\partial t}=-u \frac{\partial u}{\partial x}-v \frac{\partial u}{\partial y}-\frac{\partial p}{\partial x}+\frac{1}{\operatorname{Re}}\left(\frac{\partial^{2} u}{\partial x^{2}}+\frac{\partial^{2} u}{\partial y^{2}}\right)
\end{equation}

\begin{equation}\label{eqn:ns2}
    \frac{\partial v}{\partial t}=-u \frac{\partial v}{\partial x}-v \frac{\partial v}{\partial y}-\frac{\partial p}{\partial x}+\frac{1}{\operatorname{Re}}\left(\frac{\partial^{2} v}{\partial x^{2}}+\frac{\partial^{2} v}{\partial y^{2}}\right)
\end{equation}

\begin{equation}\label{eqn:ns3}
\frac{\partial^{2} p}{\partial x^{2}}+\frac{\partial^{2} p}{\partial y^{2}}=- \left(\left(\frac{\partial u}{\partial x}\right)^{2}+2 \frac{\partial u}{\partial y} \frac{\partial v}{\partial x}+\left(\frac{\partial v}{\partial y}\right)^{2}\right)
\end{equation}

The physics-informed loss functions are based on the residue of the PDEs, defined by the equation \ref{eqn:phyloss} in which $N_I$ denotes the number of internal nodes. The residues of the PDEs (equation \ref{eqn:residue}) consist of three sub-losses of each equation in N-S equations. Each sub-loss has sub-terms that represent the viscous term, convective term, and pressure term, respectively. 
Assume 
$\hat{u} = \frac{1}{2}(u_{i+1, j}^{n}-u_{i-1, j}^{n})$ and
$\hat{p} = \frac{1}{2}(p_{i+1, j}^{n}-p_{i-1, j}^{n})$, the three sub-terms of loss should be written as in equations \ref{eqn:lossdetail1} to \ref{eqn:lossdetail3}:

\begin{equation}\label{eqn:phyloss}
    \mathcal{L}_{phy}=\frac{1}{N_{I}}\sum_{i=1}^{N_{I}}\left\|\mathcal{R}\left(x^i,y^i\right)\right\|^{2} +\lambda_N \mathcal{L}_{Neumann}
\end{equation}

\begin{equation}\label{eqn:residue}
\mathcal{R} =\lambda_1|L_{X-momentum}| + \lambda_2|L_{Y-momentum}| + \lambda_3|L_{Continuity}|
\end{equation}

\begin{equation}\label{eqn:lossdetail1}
{L}_{X-Momentum}(\mathbf{W}, \mathbf{b})=  \frac{1}{h^2}\frac{1}{Re}\sum_{i, j}Conv2d(K, U)_{i j} + \frac{1}{h}\sum_{i,j} (u_{i,j}\hat{u}_{i,j} + v_{i,j}\hat{u}_{i,j}) -  \frac{1}{h}\hat{P}
\end{equation}

\begin{equation}\label{eqn:lossdetail2}
{L}_{Y-Momentum}(\mathbf{W}, \mathbf{b})= \frac{1}{h^2}\frac{1}{Re}\sum_{i, j}Conv2d(K, V)_{i j}+ \frac{1}{h}\sum_{i,j} (u_{i,j}\hat{v}_{i,j} + v_{i,j}\hat{v}_{i,j}) -  \frac{1}{h}\hat{P}
\end{equation}

\begin{equation}\label{eqn:lossdetail3}
 {L}_{Continuity}(\mathbf{W}, \mathbf{b})=\frac{1}{4}\sum_{i, j}Conv2d(K, P)_{i j} + {\sum_{i,j}(u_{i,j}\hat{u}_{i,j}+2\hat{u}_{i,j}\hat{v}_{i,j} + v_{i,j}\hat{v}_{i,j})}
\end{equation}

$K$ denotes the physics-informed kernels similar to kernels reported in \autocite{Sharma2018-sq} that are operated on a domain that contributes to the loss of the Laplace’s operator corresponding to the viscous terms. $Conv2d$ denotes the two dimensional convolutional operations and $h$ defines the discretization unit.\\
In this equation, another term, $\mathcal{L}_{Neumann}$ that represents the loss of any Neumann boundaries is added as a physics-driven loss. Suppose that the physics domain has the $N_X$ and $N_Y$ nodes in X and Y directions, respectively, and has $N_N$ Neumann boundary nodes in total. For the boundaries that contain Neumann B.C, equation \ref{eqn:neumann} would apply for this loss.  
\begin{equation}\label{eqn:neumann}
\mathcal{L}_{Neumann}=\frac{1}{N_{N}}\sum_{i=1}^{N_{N}}\left\|S\left(x_{0;N_Y}, y^i\right)-S\left(x_{1;N_{Y-1}},y^i\right)\right\|^{2} + \left\|S\left(x^i, y_{0;N_X}\right)-S\left(x^i,y_{1;N_{X-1}}\right)\right\|^{2}
\end{equation}

The total loss function also includes contributions from Dirichlet boundary conditions. Although the boundaries are not part of the output, they are considered in the loss function using an additional term. The data-driven loss is calculated from the mean square error between the generation and the known values on the boundaries. The boundary loss represents the distances between the values generated from the model and the actual Dirichlet B.C., which helps to reinforce the boundaries. Suppose $S = (u, v, p)$ to be physics unknown to generate and $\hat{S}$ denotes known values of the boundary at the corresponding domains and $N_b$ denotes the number of boundary nodes. For the boundaries that contain Dirichlet B.C, the equation \ref{eqn:bloss} would apply for this loss. 
\begin{equation} \label{eqn:bloss}
    \mathcal{L}_{b}=\frac{1}{N_{b}} \sum_{i=1}^{N_{b}}\left\|S\left(x_{0,N}, y^i\right)-\hat{S}\left( x_{0,N},y^i\right)\right\|^{2} + \left\|S\left(x^i, y_{0,N}\right)-\hat{S}\left(x^i,y_{0,N}\right)\right\|^{2}
\end{equation}
Finally, the total objective function comes to equation \ref{eqn:comp}:
\begin{equation}\label{eqn:comp}
\mathcal{L} =\mathcal{L}_{phy} + \lambda_b \mathcal{L}_{b}
\end{equation}

To demonstrate the facility of our method we test our model on solving the N-S equation in a two-dimensional space encompassing three physical variables, i.e., the velocity vector in the $x$ and $y$ direction and the pressure field by treating the data as a multichannel image. The data flow in this work is $D = (U, V, P; G)$, where $(U, V, P)$ are the physical variables in two-dimensional space and each represents a channel. $G$ is the optional channel corresponding to a binary geometry mask, where $1$ denotes the regions of solid boundaries considered as known boundary conditions, while $0$ stands for regular regions to be generated by the model. The input  is the high dimensional representation that denotes composite information embedded in the data structure.\\

\begin{figure}[H]
    \centering
    \includegraphics[width = \textwidth]{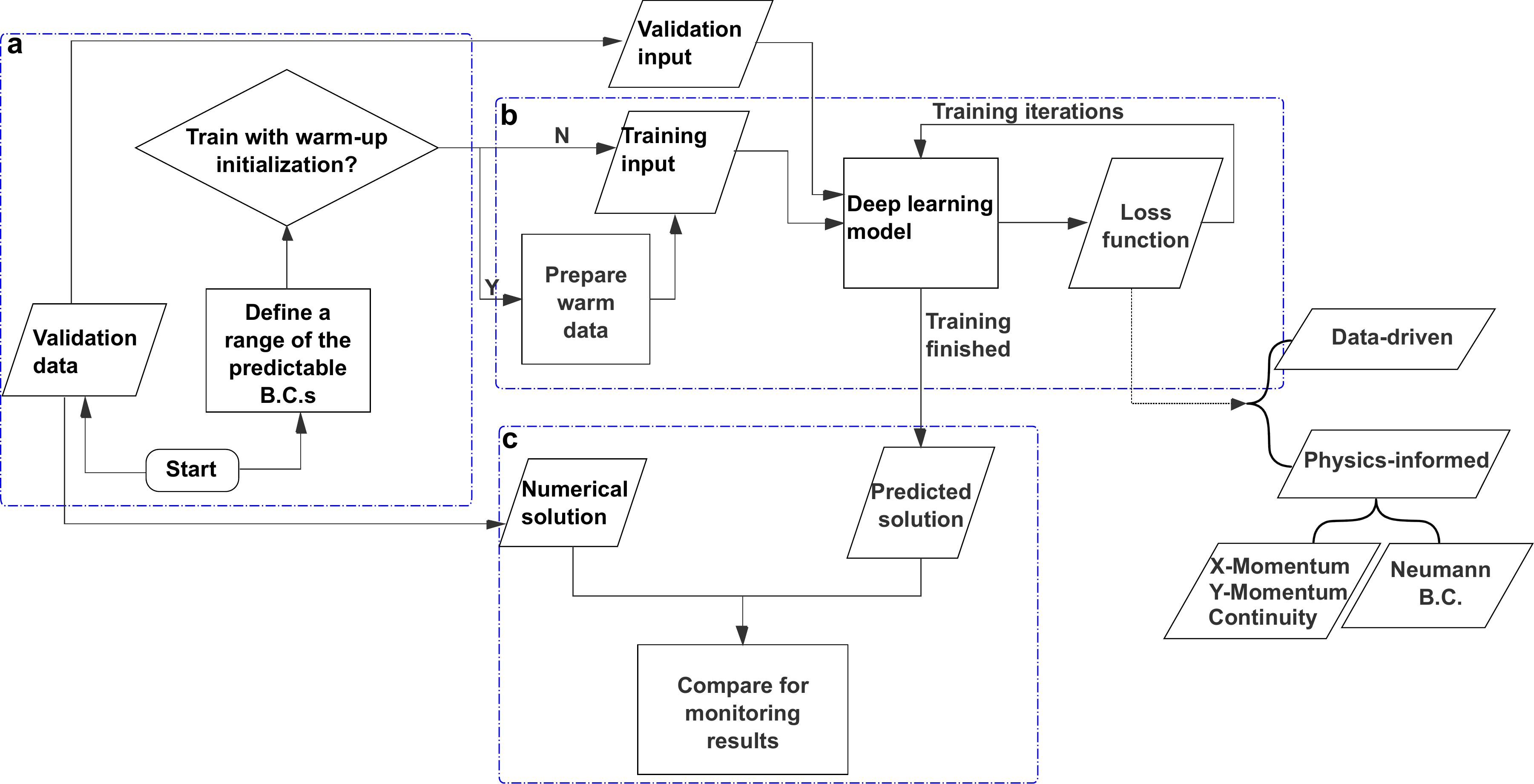}
    \caption{Workflow of the proposed method. \textbf{a}. For each training experiment, predictable B.C. is first defined, and training and validation data is prepared for training and validation purposes separately. The training data is prepared for the deep learning model with either plain initialization or warm-up initialization dependent on the problems. \textbf{b}. Training is performed with the given data and guided by the pre-defined loss functions, which consists of the hybrid of data-driven and physics-driven loss. \textbf{c}. Validation data is prepared to make the testing result given the proper initial state. A predicted solution corresponding to the given input is generated by a trained model.
}
    \label{fig:workflow}
\end{figure}

During training we follow the procedures outlined in Figure \ref{fig:workflow}. We assume a predictable range of B.C. of the model to decide the overall scale of the output from the model with  80\%-training and 20\%-testing split (Figure 1a). Using the developed deep learning framework, we train the model (Figure 1b) with the described total loss function (Equation \ref{eqn:comp}).\\ 

Using a single U-Net Model with random initialization it is hard to optimize the model to solve N-S equations. To overcome this challenge we use a stacked U-Net structure of increasing spatial resolution. Similar ideas, so-called Stack-GANs, have been used to generate realistic high-resolution images from text \autocite{Zhang2017-ty,Zhang2019-ox}. To do this, we invoke a step-by-step training process. We start by generating weakly-supervised data with pre-run iterations or coarse analytical solutions. This “warm-up” data has a minimal computational cost that requires less than 2.5 seconds to prepare for each training data. We take advantage of transfer learning at different levels during the training experiments until the model can finally achieve acceptable performance. We start by training a base model, followed by a series of succeeding models as the medium steps, and successfully acquire pre-trained models (Figure \ref{fig:modelflow}). Then we treat such trained models as a starting stage for further training steps. The base model has the capability of predicting the flow problem with a warm input. The subsequent models can consider increasing complexity including random choices of the inlet, larger domain size, geometric configurations, etc. We validate the model by comparing the numerical solutions to the predicted solutions (Figure \ref{fig:workflow} \textbf{c}). \\

\begin{figure}[H]
    \centering
    \includegraphics[width = \textwidth]{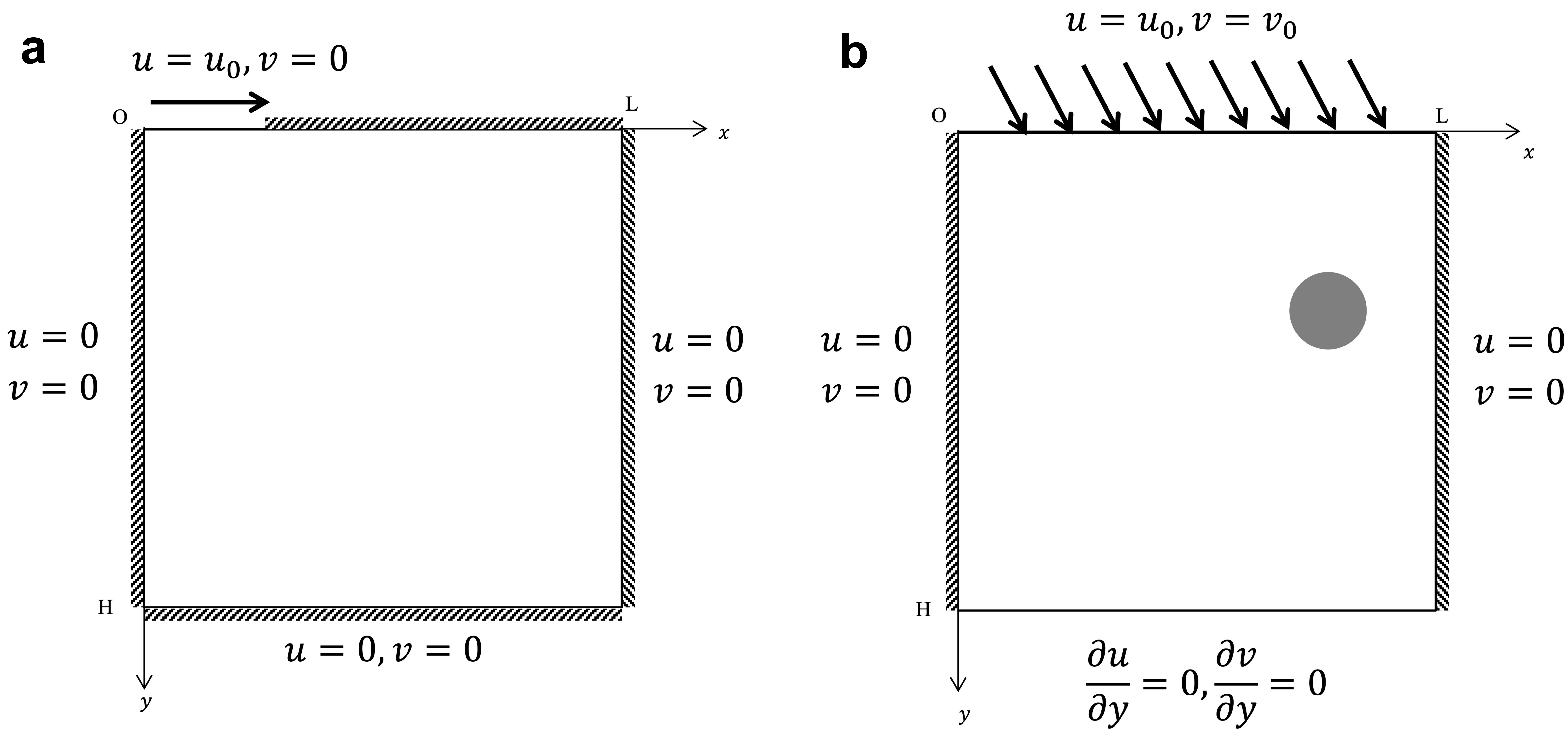}
    \caption{Schematics of the example problems. \textbf{a}. Cavity flow problem with moving lid with different lid velocities $U_0$. The moving part of the cavity lid can be changed for  model $A$; \textbf{b}. inclined flow passing over obstacles at different inlet velocities $(U_0, V_0)$. The obstacles can have different sizes and shapes.}
    \label{fig:schematics}
\end{figure}

\begin{figure}[H]
    \centering
    \includegraphics[width = \textwidth]{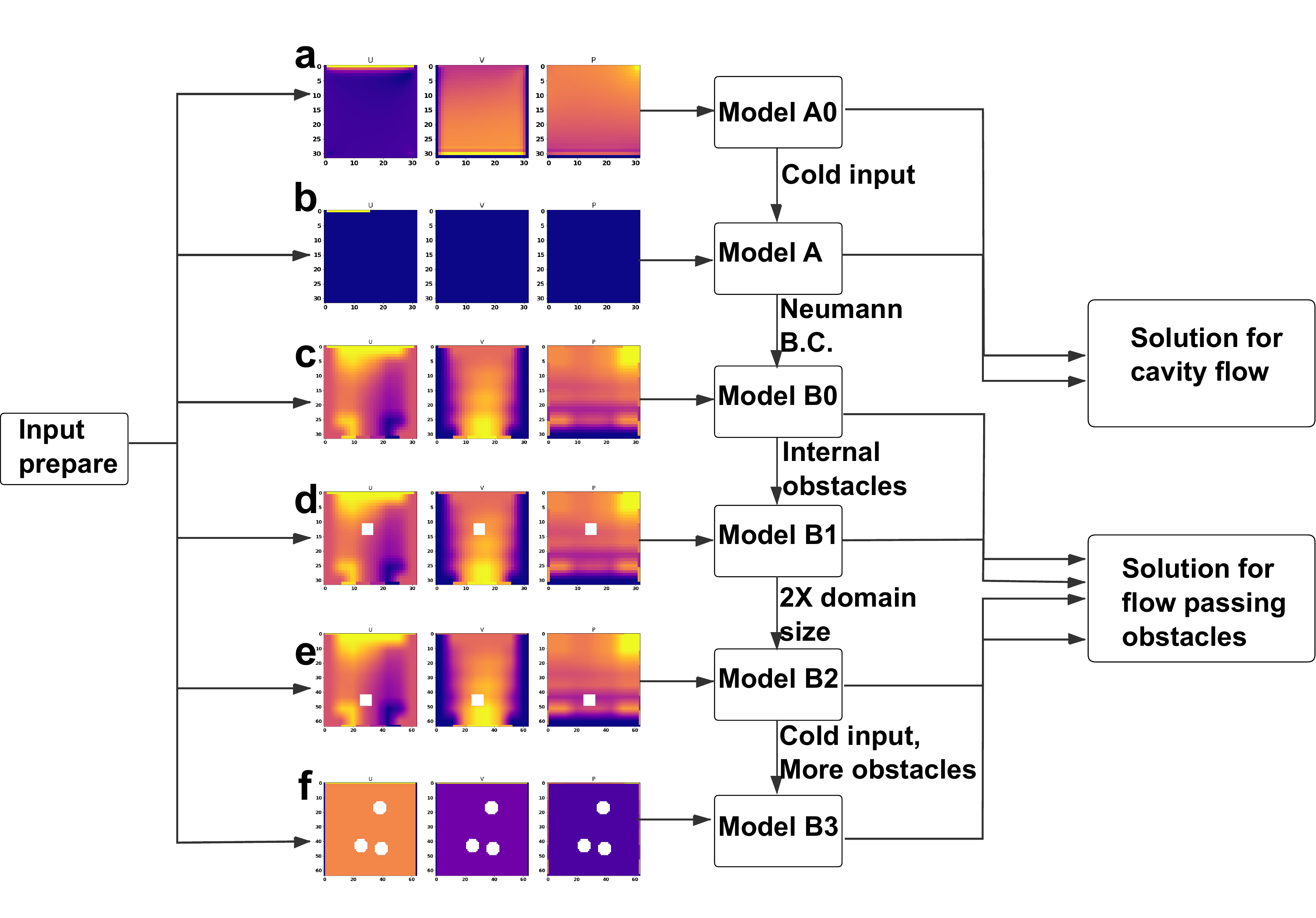}
    \caption{ Demonstration of progressive training steps for the stacked models. In the beginning, the model starts from scratch and takes warm-up initializations as the input. To reach a model with a higher level of predictability, the training is split and conducted progressively, with each step validating the corresponding result. The important new capability for the upgraded models is labeled correspondingly. About the input of each model, \textbf{a} Model $A0$ takes the input by pre-run iterations; \textbf{b} Model $A$ is trained to take the input with only the B.C.  \textbf{c}, \textbf{d} are the models that use coarse solutions, and the same warm data with a single obstacle. \textbf{e}, \textbf{f} are the models that handle larger domains. Model $B3$ can be trained to take only the B.C. and obstacle geometry as input information.
}
    \label{fig:modelflow}
\end{figure}

We start by training a model to generate a solution to the lid-driven cavity flow problem. As Figure \ref{fig:schematics} \textbf{a} shows, the enclosure is a square with a moving lid from left to right. The boundary conditions are shown in Figure\ref{fig:schematics}. The computational grid is discretized by $ 32\times 32$ grids. \\

Because we do not have a pre-trained model that allows transfer learning, the model is initialized from a $\textit{Kaiming}$ initialization \autocite{He2015-me}. The model is trained under weak supervision with low-cost computations as a warm-up (Figure \ref{fig:modelflow}\textbf{a}). The “warm-up” is not a true solution to the problem but instead is a pre-run iteration that simplifies optimization. The only cost associated with generating the training data is running an FDM iterative solver for $20$ iterations, where the Reynolds number is defined as $Re = U_0 L / \nu$, in which L is the cavity length. The lid velocities vary from 0 to 0.5, corresponding to $Re$ from 0 to 10. $Re$ can be changed by changing the lid velocity. After the first model is trained, the target output $(U, V, P)$ can be generated for different $Re$ with given pre-run steps as the input.\\

\begin{figure}[H]
    \centering
    \includegraphics[width = \textwidth]{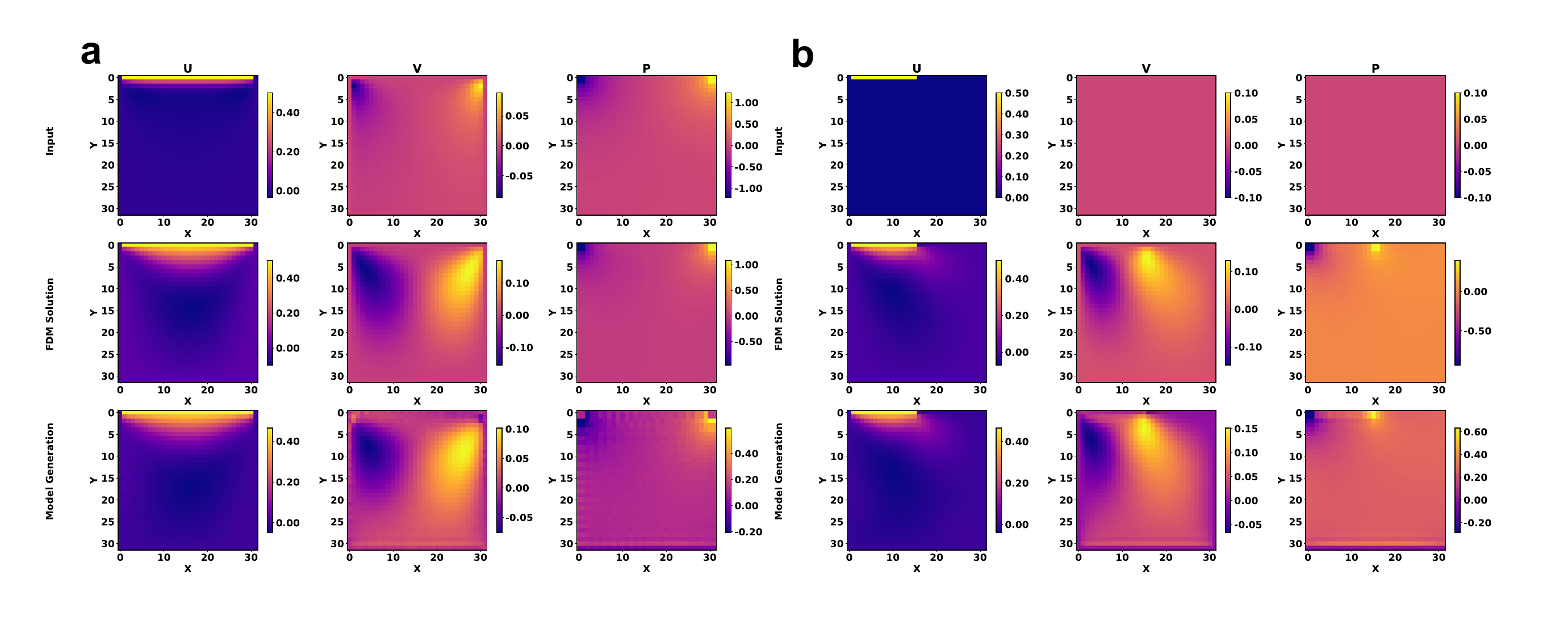}
    \caption{ Contour plots of the generation from Model $A0$ and Model $A$. a. Predicted solutions of model $A0$ tested on lid-driven flow. The lid velocity is $U_{0} = 0.5$, corresponding to $Re = 10$; b. the generation from the Model $A$ test on modified lid-driven flow. The input of the model is costless. The lid is partially moved (half-opened) and the lid velocity is $U_{0} = 0.5$, corresponding to $Re = 10$}
    \label{fig:results-cfd-noplot}
\end{figure}

To extend the generality and complexity of the base model we conducted sequential training with added constraints on each step. Due to the pre-trained base model $A0$, initialization data is not required to assist the training for model $A$, so it does not need to run a numerical solver to sample an input. Since the model is already preconditioned, the input of the second model does not require warm-up. To increase the complexity of the model we include alterations to the size and position of the upper boundary. A flowchart demonstrating the training procedures is shown in Figure \ref{fig:modelflow}\textbf{a} and \ref{fig:modelflow}\textbf{b}. The output from model $A0$ and model $A$ shows contour plots (Figure \ref{fig:results-cfd-noplot}\textbf{a}, \ref{fig:results-cfd-noplot}\textbf{b}) of the steady solution of the classic cavity lid-driven flow, and the lid-driven flow with movable lid location and sizes, respectively. The accuracy and inference speed of model $A$ and $A0$ is described in Table \ref{tab:my-table}. \\

Having proven the efficacy of our approach on the cavity-lid driven flow, we extended this concept to consider more complicated problems with inclined flow passing over obstacles at different inlet velocities We, once again, trained a base model in a weakly-supervised manner using simulated data. The base model provides a foundation for a more generalizable model that takes inputs without simulations. To demonstrate the method and test the extensibility, we design more advanced models that contain the larger domain with Neumann boundary conditions and internal obstacles. Additionally, we include a physics-informed contribution to the loss of Neumann B.C. at the domain boundaries. In this example, we train a model to generate solutions for laminar internal flow problems passing through internal obstacles. The schematic of the example is shown in Figure 3b. We consider the boundary conditions imposed at the velocity inlet and pressure outlet. The fluid flow is inclined to enter the computational domain with inlet velocity defined by $(U_0, V_0)$. Here we define the inlet velocity to be in the range of , so that the maximum magnitude of velocity is $V\leq\sqrt2/2$. The obstacles allowed in this domain do not have a characteristic length larger than the size of the domain, so the Reynolds number is small enough to be considered as laminar flow. \\

\begin{figure}[H]
    \centering
    \includegraphics[width = 0.5\textwidth]{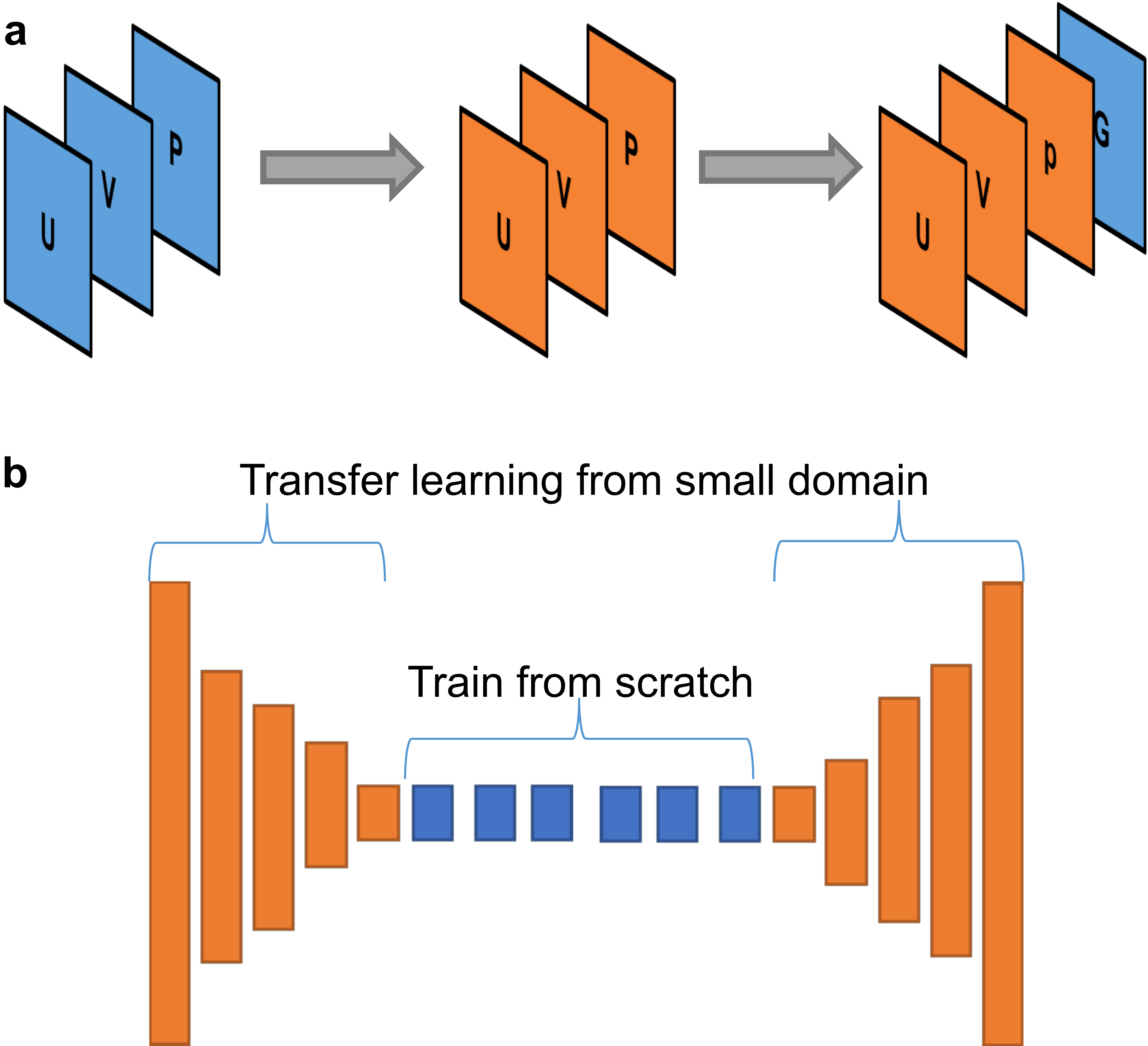}
    \caption{Demonstration of transfer learning in training when the model has a structure change. \textbf{a}: model $B0$ is developed from model $A$, the weights of the model is transferred for training model $B0$, and a geometric mask is added when training model $B1$, with the base model transfer learning from B0; \textbf{b}: during the step to improve the model from taking the input size of $ 32 \times32 $ to $ 64 \times 64 $, transfer learning is applied in a way that the inner layers are learned from scratch while the layers at the start and the end are transferred from previous model.
}
    \label{fig:transferlearning}
\end{figure}

Following the progressive training procedures, a base model (Figure \ref{fig:modelflow}\textbf{c}), Model $B0$ is developed. Similar to the cavity flow problem, this model is trained with the precondition of warm-up data, which uses the coarse computational results with the domain size of $8 \times8$ interpolated to the domain $ 32\times32 $ as the input of the model. The base model is trained to solve the internal flow problem. At this stage, the model input remains three-channel. The steady-state solutions of the domain with different values of horizontal and vertical velocities can be generated by the base model. \\

To enable approximation with internal obstacles, a geometric mask defining the obstacles is added as an additional input channel. The convolutional filter of the target model contains an extra channel. The weights for filters associated with the three channels representing the boundary are transfer-learned from the base model, while the boundary channel is optimized from scratch(Figure \ref{fig:transferlearning}\textbf{a}). We apply the transfer learning in this way to keep the training warm-started, though the structure of the model has changed due to the additional channel (Figure \ref{fig:modelflow}\textbf{d}). In this model (B1), the shape and size of the internal obstacle are fixed, and the location of this object can be various. \\

To further improve the predictability, Model $B1$ is transferred to another intermediate model, Model B2, with the increase of the computational domain size. The depth of the fully convolutional neural network changes when the input has a different size. In this case, transfer learning from Model $B1$ is performed for the layers near the start and end of the Model B2, while the inner layers require training from scratch to accommodate a domain size two times bigger (Figure \ref{fig:transferlearning}\textbf{b}).\\

\begin{figure}[H]
    \centering
    \includegraphics[width = \textwidth]{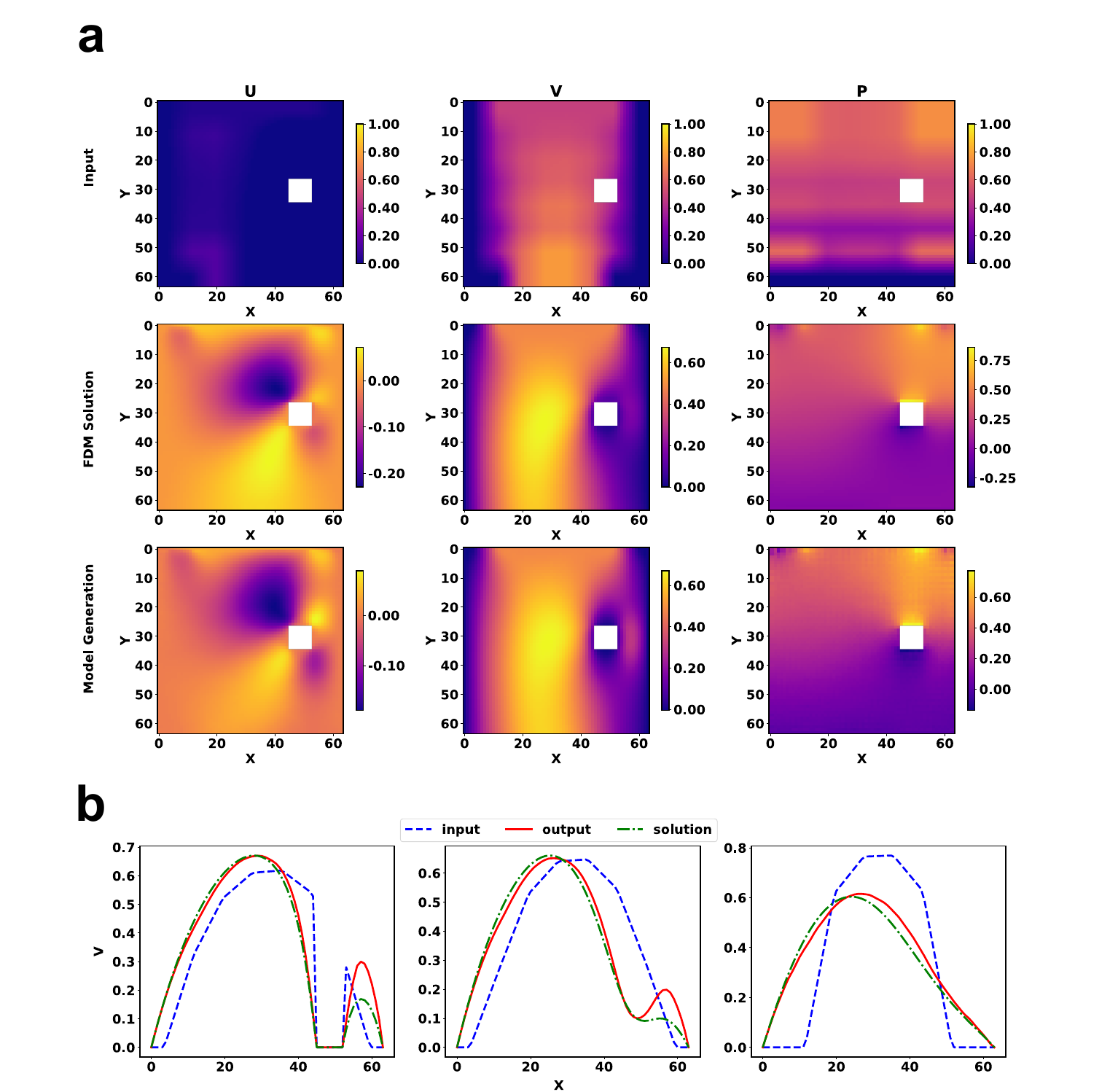}
    \caption{Generation from model $B2$ tested on an inclined velocity inlet and a single rectangle obstacle with $(U_0, V_0) = (0.05, 0.5)$. \textbf{a}. contour plots; \textbf{b}. the corresponding velocity profiles correspond to the cross-section of the center line (where the single square obstacle is located), $Y = 40$, and the outlet, respectively.}
    \label{fig:results_warm}
\end{figure}

\begin{figure}[H]
    \centering
    \includegraphics[width = \textwidth]{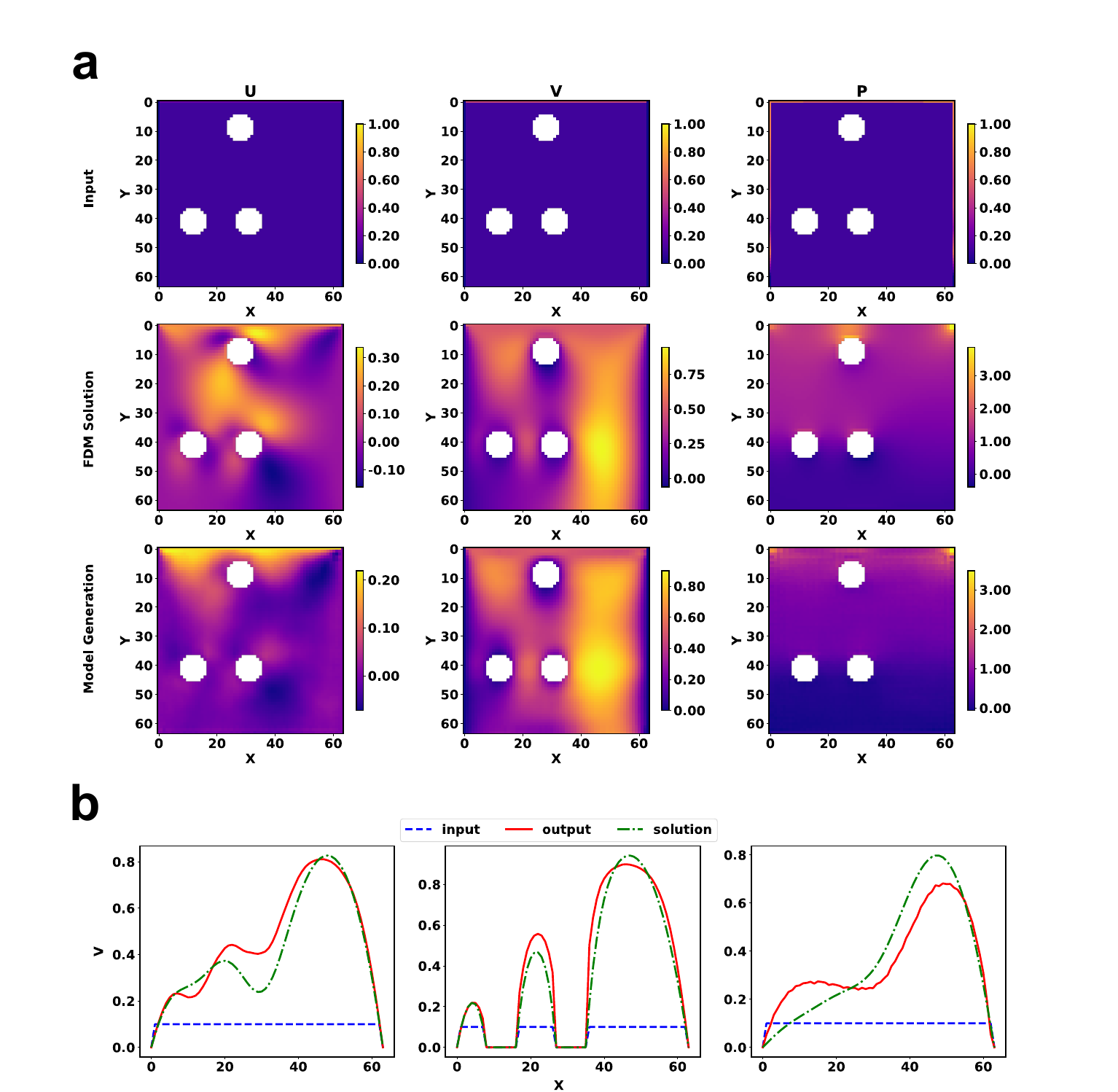}
    \caption{Generation from model $B3$ tested on an inclined velocity inlet and circle obstacles with $(U_0, V_0) = (0.2, 0.5)$. \textbf{a}. Contour plots; \textbf{b}. the corresponding velocity profiles correspond to the cross-section of the center line , $Y = 40$ (where the obstacles are located), and the outlet, respectively.
}
    \label{fig:results_cold}
\end{figure}

Finally, we conduct transfer learning and fine-tuning similar to as previously discussed to consider circle and rectangle shapes of obstacles. Without modifications to its architecture, the target model(B3) is trained and can be tested without any requirement of warm-up data, thus there is no cost for preparing the input. From the beginning of Model $B0$ to our final target, the only cost of data is the coarse CFD simulation results of pure internal flow solutions sampled from the domain $8 \times 8$ with no obstacles involved. The results from models B2 and B3 are shown in Figure \ref{fig:results_warm} and Figure \ref{fig:results_cold}, with both the contour plots (\ref{fig:results_warm} \textbf{a} and Figure \ref{fig:results_cold} \textbf{a}) and velocity profile plot at selected lines (\ref{fig:results_warm} \textbf{b},\textbf{c} and Figure \ref{fig:results_cold} \textbf{b},\textbf{c}). Testing cases for both models can take input of the inclined inlet velocity $(U_0, V_0)$ within the preset range $ (U_0, V_0) = (0, 0.5) $. Model $B2$ can generate the solutions given any location of the internal rectangle obstacle with the fixed size. Model $B3$ can handle more than one obstacle with different shapes and sizes. The accuracy of both models has been calculated as the RMSE listed in table \ref{tab:my-table}.\\

\begin{table}[H]
\centering
\caption{Summary of Example Models}
\label{tab:my-table}
\begin{tabular}{@{}l c c c c@{}}
\toprule
& \multicolumn{4}{c}{\textbf{Model}}\\
\cmidrule(l){2-5}
 &  A0 &  A &  B2 &  B3 \\ \midrule
Inlet Velocities & U = 0.5 & U = 0.5 & (U, V) = (0.05,0.5) & (U, V) = (0.2,0.5) \\
Internal obstacles & No obstacles & No obstacles & 1 square & 3 circles

 \\
Warm Input time & 0.057 seconds & \multicolumn{1}{l}{} & 0.17 seconds & \multicolumn{1}{l}{} \\
RMSE Velocity U & 0.0381 & 0.0196 & 0.0186 & 0.0836 \\
RMSE Velocity V & 0.0362 & 0.0438 & 0.0313 & 0.0766 \\
RMSE Pressure & 0.1418 & 0.2437 & 0.0619 & 0.2908 \\ \bottomrule
\end{tabular}
\end{table}


\section*{Discussion}
 
We demonstrate a highly extensible and low cost method for solving the steady N-S equations under various boundary conditions and with internal obstacles. A base model $A0$ can be successfully trained to generate solutions of the steady N-S equations for the lid-driven cavity flow problem with the help of warm-up data as the initialization and training dataset. The warm-up data that requires only pre-run steps or coarse solutions, enables the model to solve the complicated PDEs. We then train the model $A$ to solve the cavity flow problem without the warm-up data. Using the simple model that is embedded with the N-S equations as a foundation, we train a series of models ($B0$ to $B3$) to solve a different flow problem with increasing complications, considering flow passing over obstacles and a larger domain size, providing a comprehensive solution to the steady N-S equation. By transfer-learning a model trained on simple problems, an improved model can be easily trained to resolve the unseen complicated scenarios that require dealing with different boundary conditions, size of the domain, and geometric mask. Unlike a typical supervised learning, in which sufficient inputs and corresponding target outputs should be prepared as labeled data, the presented models in all training experiments do not need to fit any labeled dataset that represents the whole expected physics. Therefore, these models do not rely on large CFD datasets with existing solutions. Instead, the models take advantage of low-cost pre-inputs obtained from an FDM solver running in minutes on a desktop. 
\\

To generate solutions at different stages, the loss function requires the hybrid of physics-informed and data-driven loss to constrain the model. At the beginning of each training experiment, the generation of a model is noisy and inaccurate. The total loss of the training experiments constrains the relations of internal nodes by setting up the physics-informed loss. During the training, it is observed that if the loss solely contains the physics-informed loss, it will lose the constraint from the boundary leading to a-physical results. The data-driven loss applied on the boundary with known boundary conditions provides the necessary constraint between the internal and the boundary. By monitoring the contribution of each sub-loss, the hyperparameter for each contribution to the loss balances these constraints.\\

Our method combines the advantages of traditional CFD methods and supervised learning algorithms. The models have similar data representation as in traditional CFD. The computational domains are initiated and processed as 2D matrices, which brings good interpretability and flexibility. From training to testing, the nodal points of input retain their structure as an image, instead of isolated discrete inputs. Thanks to the similar data representation, applying a trained model only requires the direct specification of the geometric configurations and assignments of the boundary conditions. Moreover, the method is similar to a typical supervised learning problem, in which the training and testing of a model are independent. When a trained model is applied to solve N-S equations, the model in its inferencing mode only requires the computation of a forward pass with a set of fixed parameters. Given the highly parallelizable nature of the forward pass, the trained model generates the solutions in about 5 ms/iter without the need of high-performance computational resources. Following training, the model can be deployed on lightweight computational resources. We benchmark the models on a consumer desktop system with an Intel Core i5 8400 processor (6 cores and 6 threads) and achieve inference latencies within 5 ms. A visualization of the model’s instant generation can be found at \url{https://github.com/Yaling-Liu-Lab/Generate_CFD_by_DL}. This allows people with minimal deep learning knowledge to generate CFD simulation results in real time without the mathematical background or a CFD solver. One of the challenges when developing and maintaining a traditional CFD solver is the ability of running the same algorithm on different platforms with similar performance \autocite{Reguly2020-ku}. The portability of the presented model enables it to be trained on clouds with GPU resources, and then generate instant solutions on any local and portable devices equipped with only basic computing functionality. \\

While our demonstration is limited to solving the 2D-steady-state and laminar flow equations with homogeneous fluid properties this concept can be extended and adapted to other problems. For example, the model can be expanded to accommodate different geometric configurations, a wider range of predictability, different physics-informed loss, and ML-based methods for other forms of flow equations such as Stokes flow \autocite{Takbiri-Borujeni2019-lb}.

\section*{Conclusion}

A generative deep learning model implemented based on a convolutional U-Net has been developed to generate the numerical solutions for N-S equations. This model instantly produces steady-state N-S solutions on consumer computing resources. The model is facilitated by warm-up initialization and does not need computational or experimental solutions as labeled training data. The training is performed in stages adding constraints of increasing complexity, and through the inclusion of a physics-informed and data-driven loss function. By structuring the geometry as a 2D matrix similar to traditional CFD methods, the model takes the input that represents unknown variables, embedded with boundary conditions and geometric masks. The methods are validated by solving example problems including cavity flow and flow passing obstacles. The solutions generated by a series of stacked models show that the trained models can produce the steady-state solutions to a given boundary conditions for both Dirichlet and Neumann boundary and optional internal obstacles with limited cost on training data, with good predictability, extensibility, and interpretability. This base model can be extended to solve problems with larger domains, or new complexities through the addition of additional channels and subsequent optimization. Given the proper definition of the loss function, the method can also be applied to solve other PDE equations and geometric configurations. We expect that the model can be generalized to speed up the general boundary-value CFD problems and in the future, be extended to solve other fluid-structure interaction problems with minimal requirement of data.

\subsection*{Methods}

\subsubsection*{Computations in U-Net models}

The fully convolutional network contains an encoder and decoder with convolutional blocks (Figure \ref{fig:modelflow}). Each convolutional block includes convolutional layers with kernel size of $(4, 4)$ and a stride of two, batch normalization, and \textit{Leaky\_Relu} activation layer defined by equation \ref{eqn:leaky}.

\begin{equation}\label{eqn:leaky}
  \textit{Leaky\_Relu} (x) = \begin{cases} 0.2x & x \leq 0 \\ x & x>0\end{cases}   
\end{equation}

The decoder with deconvolutional layers takes the output of the preceding layer concatenated with the corresponding encoding layers. A multi-channel outcome from the decoder followed by a hyperbolic tangent $tanh$ activation layer is reconstructed to reach the original size of the input.

\subsubsection*{Training generative U-Net models}
Our models are trained for 2000 epochs on an NVIDIA 2080 Ti GPU. The model was optimized using ADAM optimizer with a learning rate of $2\times10^{-5}$. We conducted manual hyperparameter tuning including the learning rate and the multiplier of each loss, $\lambda_N, \lambda_1, \lambda_2, \lambda_3, \lambda_b$. Experiments related to the multiplier of each contribution of loss are required so that a converging loss can be reached. The multipliers should balance the contribution of each term of loss so that each of the sub-terms can reach the same magnitude.

\subsubsection*{Warm-up data preparation}
We have performed two types of warm-up data facilitating the training: pre-run solution and coarse solution. In our examples, the pre-run solution method is applied to the lid-driven cavity flow problem. This warm-up dataset is composed of 2048 samples with different lid velocities sampled from a dimensionless uniform distribution from 0 to 0.5, corresponding to $Re$ from 0 to 10. The coarse solution method is applied to the examples of flow passing obstacles. This warm-up dataset is composed of 2048 samples with the coarse computational solutions of internal flow problems without obstacles on the domain with the size of $8\times8$. The input velocities of both horizontal and vertical direction are sampled from a uniform distribution from 0 to 0.5. The coarse solutions are interpolated to the designed input size $ 32\times32 $ or  $ 64\times64 $. A geometric mask defining the locations and sizes of the obstacles is added as an additional input channel to the warm-up data.

\subsection*{Acknowledgments}
The authors acknowledge this work's support from the National Institute of Research Grant R01HL131750 and National Science Foundation grant of CBET 2039310. J.C.A acknowledge primary support from National Science Foundation under grant TRIPODS + X: RES-1839234.



\subsection*{Author Contributions} 
S.W. developed the methods, implemented the machine learning model, performed experiments, and wrote the manuscript. M.N. analyzed the fluid mechanics and wrote the manuscript. J.C.A. supervised the design of the model, guided the experiments, and wrote the manuscript. Y.L. designed the research project, analyzed results, and wrote the manuscript.

\subsection*{ORCID ID}
Shen Wang \href{https://orcid.org/0000-0002-3924-2109}{0000-0002-3924-2109}\\
Mehdi Nikfar \href{https://orcid.org/0000-0003-4575-2727}{0000-0003-4575-2727} \\
Joshua C. Agar \href{https://orcid.org/0000-0001-5411-4693}{0000-0001-5411-4693} \\
Yaling Liu \href{https://orcid.org/0000-0002-4519-3358}{0000-0002-4519-3358}

\subsection*{Code and Data availability}
The source code for implementing the method and a demonstration of the output by the model for real-time CFD simulation of an obstacle in flow can be found at: \url{  https://github.com/Yaling-Liu-Lab/Generate_CFD_by_DL}.









\subsection*{Conflicts of Interest}
The authors declare that there is no conflict of interest regarding the publication of this article.

\printbibliography

\end{document}